\documentclass[twoside,11pt]{article}

\PassOptionsToPackage{hyphens}{url}
\usepackage{jmlr2e_mod}
\usepackage[utf8]{inputenc}
\usepackage{amsmath}
\usepackage{amsfonts}
\usepackage{amssymb}
\usepackage[inline]{enumitem}
\usepackage{paralist}
\usepackage{adjustbox}
\usepackage{tikz}
\usetikzlibrary{decorations.pathmorphing}
\usetikzlibrary{arrows}
\usetikzlibrary{positioning}
\usepackage{standalone}
\usepackage{rotating}
\usepackage{tabularx}
\usepackage{booktabs}
\usepackage{multirow}
\usepackage{todonotes}
\usepackage{microtype}
\DeclareMathOperator*{\argmin}{arg\,min}
\newcommand{\Del}{\mathrm{\Delta}}

\jmlrheading{18}{2018}{1-43}{8/17}{2/18}{17-468}{Atılım Güneş Baydin, Barak~A.~Pearlmutter, Alexey~Andreyevich~Radul, and Jeffrey~Mark~Siskind}

\ShortHeadings{Automatic Differentiation in Machine Learning: a Survey}{Baydin, Pearlmutter, Radul, and Siskind}
\firstpageno{1}

\begin{document}

\title{Automatic Differentiation\\in Machine Learning: a Survey}

\author{\name Atılım Güneş Baydin \email gunes@robots.ox.ac.uk \\
       \addr Department of Engineering Science\\
       University of Oxford\\
       Oxford OX1 3PJ, United Kingdom
       \AND
       \name Barak~A.~Pearlmutter \email barak@pearlmutter.net \\
       \addr Department of Computer Science\\
       National University of Ireland Maynooth\\
       Maynooth, Co. Kildare, Ireland
       \AND
       \name Alexey~Andreyevich~Radul \email axch@mit.edu \\
       \addr Department of Brain and Cognitive Sciences\\
       Massachusetts Institute of Technology\\
       Cambridge, MA 02139, United States
       \AND
       \name Jeffrey~Mark~Siskind \email qobi@purdue.edu \\
       \addr School of Electrical and Computer Engineering\\
       Purdue University\\
       West Lafayette, IN 47907, United States}

\editor{Léon Bottou}

\maketitle

\begin{abstract}
Derivatives, mostly in the form of gradients and Hessians, are ubiquitous in machine learning. Automatic differentiation (AD), also called algorithmic differentiation or simply ``autodiff'', is a family of techniques similar to but more general than backpropagation for efficiently and accurately evaluating derivatives of numeric functions expressed as computer programs. AD is a small but established field with applications in areas including computational fluid dynamics, atmospheric sciences, and engineering design optimization. Until very recently, the fields of machine learning and AD have largely been unaware of each other and, in some cases, have independently discovered each other's results. Despite its relevance, general-purpose AD has been missing from the machine learning toolbox, a situation slowly changing with its ongoing adoption under the names ``dynamic computational graphs'' and ``differentiable programming''. We survey the intersection of AD and machine learning, cover applications where AD has direct relevance, and address the main implementation techniques. By precisely defining the main differentiation techniques and their interrelationships, we aim to bring clarity to the usage of the terms ``autodiff'', ``automatic differentiation'', and ``symbolic differentiation'' as these are encountered more and more in machine learning settings.

\end{abstract}


\section{Introduction}

Methods for the computation of derivatives in computer programs can be classified into four categories:
\begin{inparaenum}[(1)]
    \item manually working out derivatives and coding them;
    \item \emph{numerical differentiation} using finite difference approximations;
    \item \emph{symbolic differentiation} using expression manipulation in computer algebra systems such as Mathematica, Maxima, and Maple; and
    \item \emph{automatic differentiation}, also called \emph{algorithmic differentiation}, which is the subject matter of this paper.
\end{inparaenum}

Conventionally, many methods in machine learning have required the evaluation of derivatives and most of the traditional learning algorithms have relied on the computation of gradients and Hessians of an objective function \citep{Sra2011}. When introducing new models, machine learning researchers have spent considerable effort on the manual derivation of analytical derivatives to subsequently plug these into standard optimization procedures such as L-BFGS \citep{Zhu1997} or stochastic gradient descent \citep{Bottou1998}. Manual differentiation is time consuming and prone to error. Of the other alternatives, numerical differentiation is simple to implement but can be highly inaccurate due to round-off and truncation errors \citep{Jerrell1997}; more importantly, it scales poorly for gradients, rendering it inappropriate for machine learning where gradients with respect to millions of parameters are commonly needed. Symbolic differentiation addresses the weaknesses of both the manual and numerical methods, but often results in complex and cryptic expressions plagued with the problem of ``expression swell'' \citep{Corliss1988}. Furthermore, manual and symbolic methods require models to be defined as closed-form expressions, ruling out or severely limiting algorithmic control flow and expressivity.

We are concerned with the powerful fourth technique, automatic differentiation (AD). AD performs a non-standard interpretation of a given computer program by replacing the domain of the variables to incorporate derivative values and redefining the semantics of the operators to propagate derivatives per the chain rule of differential calculus. Despite its widespread use in other fields, general-purpose AD has been underused by the machine learning community until very recently.\footnote{See, e.g., \url{https://justindomke.wordpress.com/2009/02/17/automatic-differentiation-the-most-criminally-underused-tool-in-the-potential-machine-learning-toolbox/}} Following the emergence of deep learning \citep{lecun2015deep,goodfellow2016deep} as the state-of-the-art in many machine learning tasks and the modern workflow based on rapid prototyping and code reuse in frameworks such as Theano \citep{Bastien2012}, Torch \citep{collobert2011torch7}, and TensorFlow \citep{abadi2016tensorflow}, the situation is slowly changing where projects such as autograd\footnote{\url{https://github.com/HIPS/autograd}} \citep{maclaurin2016modeling}, Chainer\footnote{\url{https://chainer.org/}} \citep{tokui2015chainer}, and PyTorch\footnote{\url{http://pytorch.org/}} \citep{paszke2017automatic} are leading the way in bringing general-purpose AD to the mainstream.

The term ``automatic'' in AD can be a source of confusion, causing machine learning practitioners to put the label ``automatic differentiation'', or just ``autodiff'', on any method or tool that does not involve manual differentiation, without giving due attention to the underlying mechanism. We would like to stress that AD as a technical term refers to a specific family of techniques that compute derivatives through accumulation of values during code execution to generate numerical derivative evaluations rather than derivative expressions. This allows accurate evaluation of derivatives at machine precision with only a small constant factor of overhead and ideal asymptotic efficiency. In contrast with the effort involved in arranging code as closed-form expressions under the syntactic and semantic constraints of symbolic differentiation, AD can be applied to regular code with minimal change, allowing branching, loops, and recursion. Because of this generality, AD has been applied to computer simulations in industry and academia and found applications in fields including engineering design optimization \citep{forth2002aerofoil,casanova2002application}, computational fluid dynamics \citep{Muller2005,thomas2006using,Bischof2006}, physical modeling \citep{Ekstrom2010}, optimal control \citep{Walther2007}, structural mechanics \citep{haase2002optimal}, atmospheric sciences \citep{Carmichael1997,Charpentier2000}, and computational finance \citep{Bischof2002,Capriotti2011}.

\begin{figure}
  \centering
  \trimbox{0cm -0.4cm}{\resizebox{0.75\textwidth}{!}{\includegraphics{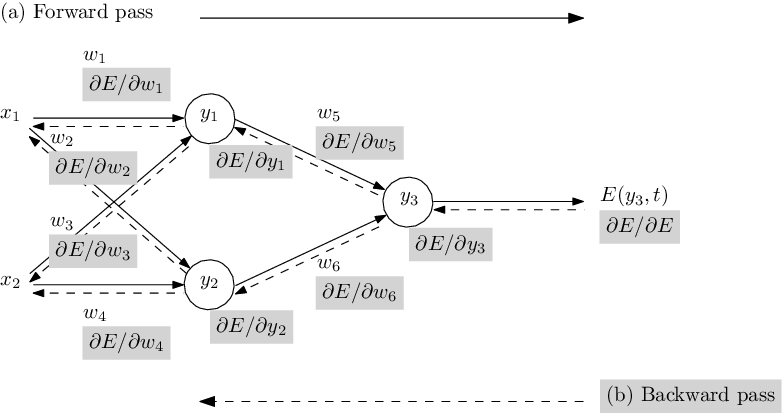}}}
  \caption{Overview of backpropagation. (a) Training inputs $x_i$ are fed forward, generating corresponding activations $y_i$. An error $E$ between the actual output $y_3$ and the target output $t$ is computed. (b) The error adjoint is propagated backward, giving the gradient with respect to the weights $\nabla_{w_i}E = \left(\frac{\partial E}{\partial w_1},\dots,\frac{\partial E}{\partial w_6}\right)$, which is subsequently used in a gradient-descent procedure. The gradient with respect to inputs $\nabla_{x_i}E$ can be also computed in the same backward pass.}
  \label{FigureBackpropagation}
\end{figure}

In machine learning, a specialized counterpart of AD known as the backpropagation algorithm has been the mainstay for training neural networks, with a colorful history of having been reinvented at various times by independent researchers \citep{Griewank2012,schmidhuber2015deep}. It has been one of the most studied and used training algorithms since the day it became popular mainly through the work of \citet{rumelhart1986learning}. In simplest terms, backpropagation models learning as gradient descent in neural network weight space, looking for the minima of an objective function. The required gradient is obtained by the backward propagation of the sensitivity of the objective value at the output (Figure~\ref{FigureBackpropagation}), utilizing the chain rule to compute partial derivatives of the objective with respect to each weight. The resulting algorithm is essentially equivalent to transforming the network evaluation function composed with the objective function under reverse mode AD, which, as we shall see, actually generalizes the backpropagation idea. Thus, a modest understanding of the mathematics underlying backpropagation provides one with sufficient background for grasping AD techniques.

In this paper we review AD from a machine learning perspective, covering its origins, applications in machine learning, and methods of implementation. Along the way, we also aim to dispel some misconceptions that we believe have impeded wider recognition of AD by the machine learning community. In Section~\ref{SectionWhatADIsNot} we start by explicating how AD differs from numerical and symbolic differentiation. Section~\ref{SectionPreliminaries} gives an introduction to the AD technique and its forward and reverse accumulation modes. Section~\ref{SectionDerivativesAndMachineLearning} discusses the role of derivatives in machine learning and examines cases where AD has relevance. Section~\ref{SectionImplementations} covers various implementation approaches and general-purpose AD tools, followed by Section~\ref{SectionConclusions} where we discuss future directions.

\section{What AD Is Not}
\label{SectionWhatADIsNot}

Without proper introduction, one might assume that AD is either a type of numerical or symbolic differentiation. Confusion can arise because AD does in fact provide numerical values of derivatives (as opposed to derivative expressions) and it does so by using symbolic rules of differentiation (but keeping track of derivative values as opposed to the resulting expressions), giving it a two-sided nature that is partly symbolic and partly numerical \citep{Griewank2003}. We start by emphasizing how AD is different from, and in several aspects superior to, these two commonly encountered techniques of computing derivatives.

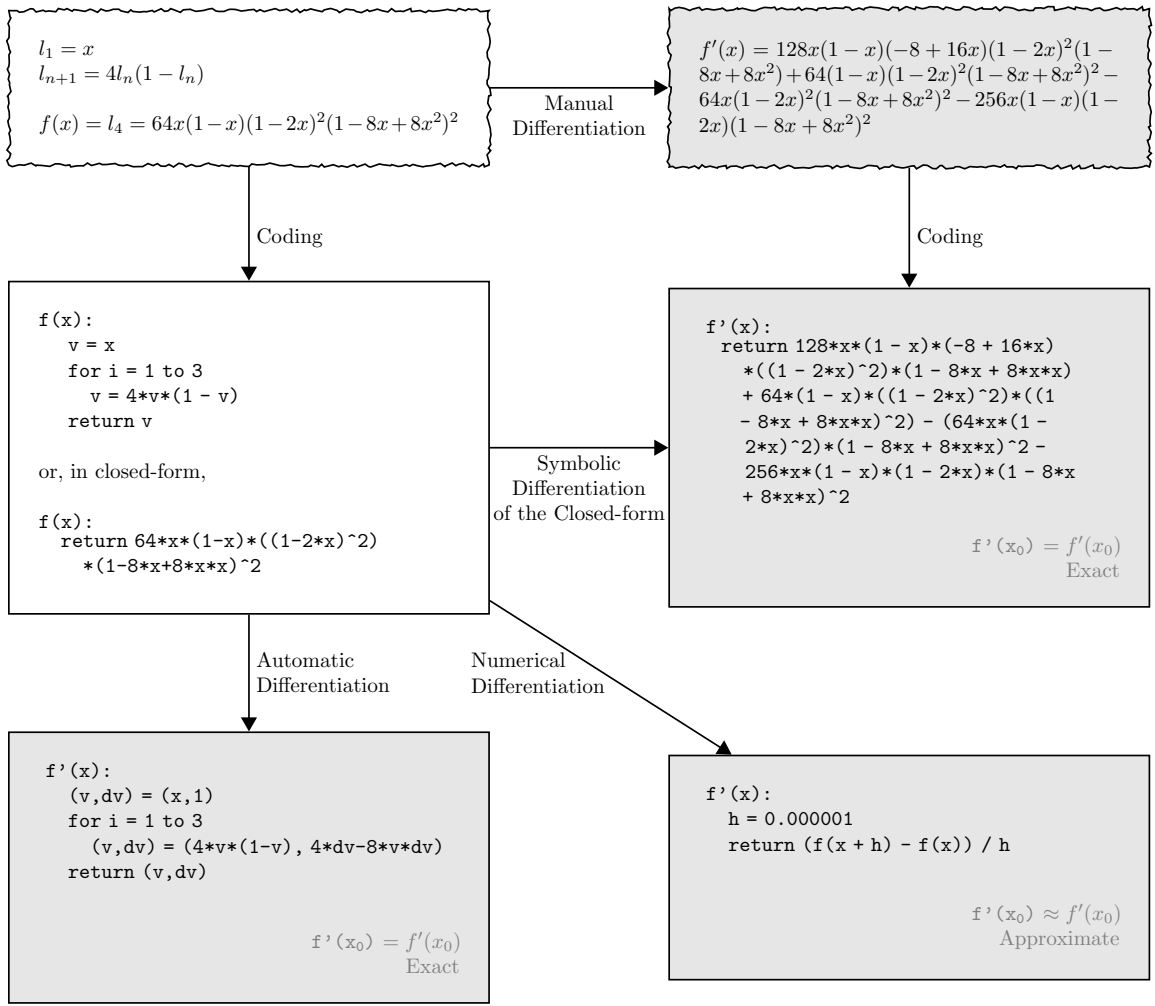
\begin{figure*}
  \centering
  \trimbox{0cm -0.4cm}{\resizebox{\textwidth}{!}{\small\begin{tikzpicture}[
	pencildraw/.style={
		decorate, 
		decoration={random steps,segment length=2pt,amplitude=1pt}
	}]
	\tikzstyle{paperbox} = [pencildraw,draw,thick,fill=white,text width=7cm,inner sep=5mm]
	\tikzstyle{codebox} = [draw,thick,text width=7cm,inner sep=5mm]
	\tikzstyle{edge} = [->,>=triangle 60,thick]
	
    	\node[paperbox] (a) at (-5.5,-12.5) {
    	$l_1=x$\\
    	$l_{n+1}=4l_n(1-l_n)$\\
    	\vspace{4mm}
    	$f(x)=l_4=64x(1 - x)(1 - 2 x)^2 (1 - 8 x + 8 x^2)^2$
    	};
    	\node[paperbox,fill=gray!20] (b) at (5.5,-12.5) {$f'(x)=128x(1 - x)(-8 + 16 x)(1 - 2 x)^2(1 - 8 x + 8 x^2) + 64 (1 - x)(1 - 2 x)^2  (1 - 8 x + 8 x^2)^2 - 64x(1 - 2 x)^2 (1 - 8 x + 8 x^2)^2 - 256x(1 - x)(1 - 2 x)(1 - 8 x + 8 x^2)^2$};
    	\node[codebox,fill=white] (c) at (-5.5,-18.5) {\texttt{\parbox{7cm}{
		f(x):\\
    		\hphantom{tt} v = x\\
     		\hphantom{tt} for i = 1 to 3\\
        	\hphantom{tttt} v = 4*v*(1 - v)\\
    		\hphantom{tt} return v\\
		\hphantom{t}\\
    		\textrm{or, in closed-form,}\\
    		\hphantom{t}\\
    		f(x):\\
    		\hphantom{tt}\parbox{6cm}{\hangindent=0.4cm \hangafter=1 return 64*x*(1-x)*((1-2*x)\^{}2)\\ *(1-8*x+8*x*x)\^{}2}\\
	}}};
    	\node[codebox,black,fill=gray!20] (d) at (5.5,-18.5) {\color{black}\texttt{\parbox{7cm}{\textbf{
    		f'(x):\\
		\hphantom{tt}\parbox{6cm}{\hangindent=0.4cm \hangafter=1 return 128*x*(1 - x)*(-8 + 16*x)\\ *((1 - 2*x)\^{}2)*(1 - 8*x + 8*x*x)\\+ 64*(1 - x)*((1 - 2*x)\^{}2)*((1 - 8*x + 8*x*x)\^{}2) - (64*x*(1 - 2*x)\^{}2)*(1 - 8*x + 8*x*x)\^{}2 - 256*x*(1 - x)*(1 - 2*x)*(1 - 8*x + 8*x*x)\^{}2}\\}
		\flushright \color{gray} f'($\mathtt{x_0}$) $= f'(x_0)$\\\textrm{Exact}
	}}};
    	\node[codebox,black,fill=gray!20] (e) at (-5.5,-25.5) {\color{black}\texttt{\parbox{7cm}{\textbf{
		f'(x):\\
    		\hphantom{tt} (v,dv) = (x,1)\\
    		\hphantom{tt} for i = 1 to 3\\
        	\hphantom{tttt} (v,dv) = (4*v*(1-v), 4*dv-8*v*dv)\\
    		\hphantom{tt} return (v,dv)\\}
    		\flushright \color{gray} f'($\mathtt{x_0}$) $= f'(x_0)$\\\textrm{Exact}
	}}};
    	\node[codebox,black,fill=gray!20] (f) at (5.5,-25.5) {\color{black}\texttt{\parbox{7cm}{\textbf{
    		f'(x):\\
        \hphantom{tt} h = 0.000001\\
		\hphantom{tt} return (f(x + h) - f(x)) / h\\}
		\flushright \color{gray} f'($\mathtt{x_0}$) $\approx f'(x_0)$\\\textrm{Approximate}
	}}};
	
	\draw (a) edge [edge] (b);
    	\node[align=center,below] at (0,-12.5) {Manual\\Differentiation};
	
	\draw (c) edge [edge] (d);
    	\node[align=center,below] at (0,-18.5) {Symbolic\\Differentiation\\of the Closed-form};
    	
	\draw (a) edge [edge] (c);
	\node[align=left,right] at (-5.5,-15) {Coding};
    	
	\draw (b) edge [edge] (d);
	\node[align=left,right] at (5.5,-15) {Coding};
    	
	\draw (c) edge [edge] (f);
	\node[align=left,left] at (0.55,-22.25) {Numerical\\Differentiation};
    	
    	\draw (c) edge [edge] (e);
    	\node[align=left,right] at (-5.5,-22.25) {Automatic\\Differentiation};
    	
\end{tikzpicture}}}
  \caption{The range of approaches for differentiating mathematical expressions and computer code, looking at the example of a truncated logistic map (upper left). Symbolic differentiation (center right) gives exact results but requires closed-form input and suffers from expression swell; numerical differentiation (lower right) has problems of accuracy due to round-off and truncation errors; automatic differentiation (lower left) is as accurate as symbolic differentiation with only a constant factor of overhead and support for control flow.}
  \label{FigureDifferentiation}
\end{figure*}

\subsection{AD Is Not Numerical Differentiation}

Numerical differentiation is the finite difference approximation of derivatives using values of the original function evaluated at some sample points \citep{Burden2001} (Figure~\ref{FigureDifferentiation}, lower right). In its simplest form, it is based on the limit definition of a derivative. For example, for a multivariate function $f:\mathbb{R}^n \to \mathbb{R}$, one can approximate the gradient $\nabla f=\left(\frac{\partial f}{\partial x_1},\dots,\frac{\partial f}{\partial x_n}\right)$ using
\begin{equation}
  \label{EquationForwardDifference}
  \frac{\partial f(\mathbf{x})}{\partial x_i} \approx \frac{f(\mathbf{x} + h \mathbf{e}_i) - f(\mathbf{x})}{h}\;,
\end{equation}
where $\mathbf{e}_i$ is the $i$-th unit vector and $h > 0$ is a small step size. This has the advantage of being uncomplicated to implement, but the disadvantages of performing $O(n)$ evaluations of $f$ for a gradient in $n$ dimensions and requiring careful consideration in selecting the step size $h$.

\begin{figure*}
  \centering
  \resizebox{0.82\textwidth}{!}{\small\input{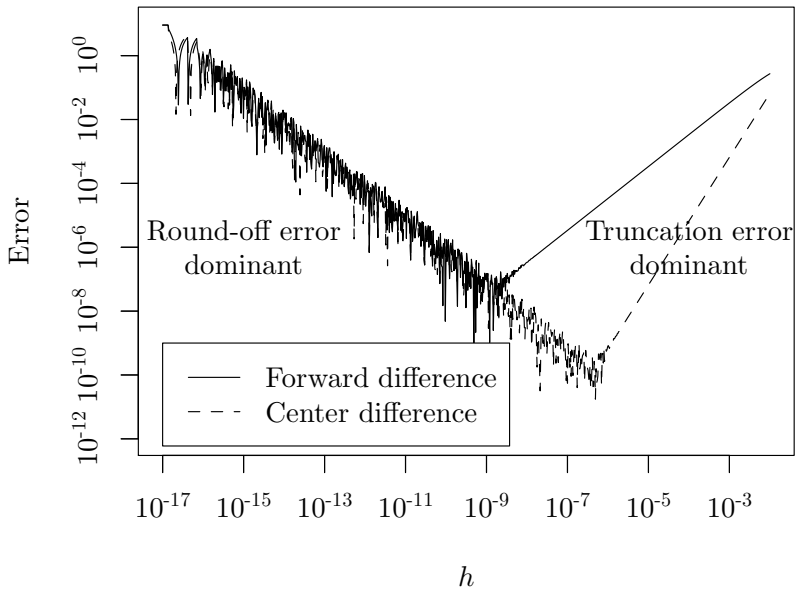}}
  \caption{Error in the forward (Eq.~\ref{EquationForwardDifference}) and center difference (Eq.~\ref{EquationCenterDifference}) approximations as a function of step size $h$, for the derivative of the truncated logistic map ${f(x)=64x(1 - x)(1 - 2 x)^2 (1 - 8 x + 8 x^2)^2}$. Plotted errors are computed using ${E_{\mathrm{forward}}(h,x_0)=\left|\frac{f(x_0+h)-f(x_0)}{h} - \frac{d}{dx}f(x)\big|_{x_0}\right|}$ and ${E_{\mathrm{center}}(h,x_0)=\left|\frac{f(x_0+h)-f(x_0-h)}{2h} - \frac{d}{dx}f(x)\big|_{x_0}\right|}$ at $x_0=0.2$\;.}
  \label{FigureApproximationError}
\end{figure*}

Numerical approximations of derivatives are inherently ill-conditioned and unstable,\footnote{Using the limit definition of the derivative for finite difference approximation commits both cardinal sins of numerical analysis: \emph{``thou shalt not add small numbers to big numbers''}, and \emph{``thou shalt not subtract numbers which are approximately equal''}.} with the exception of complex variable methods that are applicable to a limited set of holomorphic functions \citep{Fornberg1981}. This is due to the introduction of truncation\footnote{Truncation error is the error of approximation, or inaccuracy, one gets from $h$ not actually being zero. It is proportional to a power of $h$.} and round-off\footnote{Round-off error is the inaccuracy one gets from valuable low-order bits of the final answer having to compete for machine-word space with high-order bits of $f(\mathbf{x} + h \mathbf{e}_i)$ and $f(\mathbf{x})$ (Eq.~\ref{EquationForwardDifference}), which the computer has to store just until they cancel in the subtraction at the end. Round-off error is inversely proportional to a power of $h$.} errors inflicted by the limited precision of computations and the chosen value of the step size $h$. Truncation error tends to zero as $h \to 0$. However, as $h$ is decreased, round-off error increases and becomes dominant (Figure~\ref{FigureApproximationError}).

Various techniques have been developed to mitigate approximation errors in numerical differentiation, such as using a center difference approximation
\begin{equation}
  \label{EquationCenterDifference}
  \frac{\partial f(\mathbf{x})}{\partial x_i} = \frac{f(\mathbf{x} + h \mathbf{e}_i) - f(\mathbf{x} - h \mathbf{e}_i)}{2h} + O(h^{2})\;,
\end{equation}
where the first-order errors cancel and one effectively moves the truncation error from first-order to second-order in $h$.\footnote{This does not avoid either of the cardinal sins, and is still highly inaccurate due to truncation.} For the one-dimensional case, it is just as costly to compute the forward difference (Eq.~\ref{EquationForwardDifference}) and the center difference (Eq.~\ref{EquationCenterDifference}), requiring only two evaluations of $f$. However, with increasing dimensionality, a trade-off between accuracy and performance is faced, where computing a Jacobian matrix of a function $f: \mathbb{R}^n \to \mathbb{R}^m$ requires $2mn$ evaluations.

Other techniques for improving numerical differentiation, including higher-order finite differences, Richardson extrapolation to the limit \citep{Brezinski1991}, and differential quadrature methods using weighted sums \citep{Bert1996}, have increased computational complexity, do not completely eliminate approximation errors, and remain highly susceptible to floating point truncation.

The $O(n)$ complexity of numerical differentiation for a gradient in $n$ dimensions is the main obstacle to its usefulness in machine learning, where $n$ can be as large as millions or billions in state-of-the-art deep learning models \citep{shazeer2017outrageously}. In contrast, approximation errors would be tolerated in a deep learning setting thanks to the well-documented error resiliency of neural network architectures \citep{gupta2015deep}.

\subsection{AD Is Not Symbolic Differentiation}

Symbolic differentiation is the automatic manipulation of expressions for obtaining derivative expressions \citep{Grabmeier2003} (Figure~\ref{FigureDifferentiation}, center right), carried out by applying transformations representing rules of differentiation such as
\begin{equation}
\begin{aligned}
\frac{d}{dx} \left(f(x) + g(x)\right) &\leadsto \frac{d}{dx} f(x) + \frac{d}{dx} g(x)\\
\frac{d}{dx} \left(f(x)\,g(x)\right) &\leadsto \left(\frac{d}{dx} f(x)\right) g(x) + f(x) \left(\frac{d}{dx} g(x)\right)\; .
\end{aligned}
\label{EquationMultiplicationRule}
\end{equation}

When formulae are represented as data structures, symbolically differentiating an expression tree is a perfectly mechanistic process, considered subject to mechanical automation even at the very inception of calculus \citep{Leibniz1685}. This is realized in modern computer algebra systems such as Mathematica, Maxima, and Maple and machine learning frameworks such as Theano.

In optimization, symbolic derivatives can give valuable insight into the structure of the problem domain and, in some cases, produce analytical solutions of extrema (e.g., solving for $\frac{d}{dx}f(x)=0$) that can eliminate the need for derivative calculation altogether. On the other hand, symbolic derivatives do not lend themselves to efficient runtime calculation of derivative values, as they can get exponentially larger than the expression whose derivative they represent.

Consider a function $h(x)=f(x)g(x)$ and the multiplication rule in Eq.~\ref{EquationMultiplicationRule}. Since $h$ is a product, $h(x)$ and $\frac{d}{dx}h(x)$ have some common components, namely $f(x)$ and $g(x)$. Note also that on the right hand side, $f(x)$ and $\frac{d}{dx}f(x)$ appear separately. If we just proceeded to symbolically differentiate $f(x)$ and plugged its derivative into the appropriate place, we would have nested duplications of any computation that appears in common between $f(x)$ and $\frac{d}{dx}f(x)$. Hence, careless symbolic differentiation can easily produce exponentially large symbolic expressions which take correspondingly long to evaluate. This problem is known as \emph{expression swell} (Table~\ref{TableExpressionSwell}).

\begin{table}
  \centering
  \renewcommand{\arraystretch}{1.2}
  \caption{Iterations of the logistic map $l_{n+1}=4l_n (1-l_n)$, $l_1=x$ and the corresponding derivatives of $l_n$ with respect to $x$, illustrating expression swell.}
  \label{TableExpressionSwell}
  {\small
  \begin{tabularx}{\columnwidth}{@{}lp{2.8cm}XX@{}}
    \toprule
    $n$ & $l_n$ & $\frac{d}{dx}l_n$ & $\frac{d}{dx}l_n$ (Simplified form)\\
    \addlinespace
    \midrule
    1 & $x$ & $1$ & $1$\\
    \addlinespace
    2 & $4x(1 - x)$ & $4(1 - x) -4x$ & $4 - 8x$\\
    \addlinespace
    3 & $16x(1 - x)(1 - 2 x)^2$ & $16(1 - x)(1 - 2 x)^2 - 16x(1 - 2 x)^2 - 64x(1 - x)(1 - 2 x)$ & $16 (1 - 10 x + 24 x^2 - 16 x^3)$\\
    \addlinespace
    4 & $64x(1 - x)(1 - 2 x)^2$ $(1 - 8 x + 8 x^2)^2$ & $128x(1 - x)(-8 + 16 x)(1 - 2 x)^2 (1 - 8 x + 8 x^2) + 64 (1 - x)(1 - 2 x)^2  (1 - 8 x + 8 x^2)^2 - 64x(1 - 2 x)^2 (1 - 8 x + 8 x^2)^2 - 256x(1 - x)(1 - 2 x)(1 - 8 x + 8 x^2)^2$ & $64 (1 - 42 x + 504 x^2 - 2640 x^3 + 7040 x^4 - 9984 x^5 + 7168 x^6 - 2048 x^7)$\\
    \bottomrule
  \end{tabularx}}
\end{table}

When we are concerned with the accurate numerical evaluation of derivatives and not so much with their actual symbolic form, it is in principle possible to significantly simplify computations by storing only the values of intermediate sub-expressions in memory. Moreover, for further efficiency, we can interleave as much as possible the differentiation and simplification steps. This interleaving idea forms the basis of AD and provides an account of its simplest form: \emph{apply symbolic differentiation at the elementary operation level and keep intermediate numerical results, in lockstep with the evaluation of the main function.} This is AD in the forward accumulation mode, which we shall introduce in the following section.

\section{AD and Its Main Modes}
\label{SectionPreliminaries}

AD can be thought of as performing a non-standard interpretation of a computer program where this interpretation involves augmenting the standard computation with the calculation of various derivatives. All numerical computations are ultimately compositions of a finite set of elementary operations for which derivatives are known \citep{Verma2000,Griewank2008}, and combining the derivatives of the constituent operations through the chain rule gives the derivative of the overall composition. Usually these elementary operations include the binary arithmetic operations, the unary sign switch, and transcendental functions such as the exponential, the logarithm, and the trigonometric functions.

\begin{figure}
  \centering
  \trimbox{0cm -0.5cm}{\resizebox{0.8\textwidth}{!}{\normalsize\begin{tikzpicture}[]
	
	\tikzstyle{vnode} = [circle,draw,thick,fill=white,minimum size=9mm]
	\tikzstyle{vedge} = [->,>=latex,thick]
	
	\node[vnode] (v-1) at (-8.5,0.5) {$v_{-1}$};
	\node[vnode] (v0) at (-8.5,-2.5) {$v_0$};
	\node[vnode] (v1) at (-6,0.5) {$v_1$};
	\node[vnode] (v2) at (-6,-1) {$v_2$};
	\node[vnode] (v3) at (-3.5,-2.5) {$v_3$};
	\node[vnode] (v4) at (-3.5,0.5) {$v_4$};
	\node[vnode] (v5) at (-1,-1) {$v_5$};
	
	\node[] (x1) at (-11,0.5) {$x_1$};
	\node[] (x2) at (-11,-2.5) {$x_2$};
	\node[] (f) at (1.5,-1) {$f(x_1,x_2)$};
	
	\draw (v-1) edge [vedge] (v1);
	\draw (v-1) edge [vedge] (v2);
	\draw (v0) edge [vedge] (v2);
	\draw (v0) edge [vedge] (v3);
	\draw (v1) edge [vedge] (v4);
	\draw (v2) edge [vedge] (v4);
	\draw (v3) edge [vedge] (v5);
	\draw (v4) edge [vedge] (v5);
	
	\draw (x1) edge [vedge] (v-1);
	\draw (x2) edge [vedge] (v0);
	\draw (v5) edge [vedge] (f);

\end{tikzpicture}}}
  \caption{Computational graph of the example $f(x_1, x_2) = \ln(x_1) + x_1 x_2 - \sin(x_2)$. See the primal trace in Tables \ref{TableForwardADExample} or \ref{TableReverseADExample} for the definitions of the intermediate variables $v_{-1} \dots v_5$\;.}
  \label{FigureComputationalGraph}
\end{figure}
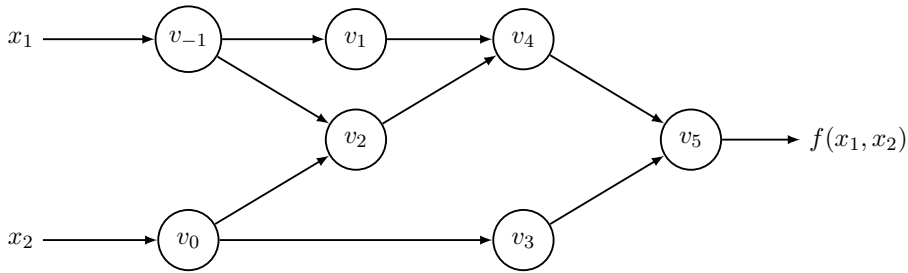

On the left hand side of Table~\ref{TableForwardADExample} we see the representation of the computation $y = f(x_1, x_2) = \ln(x_1) + x_1 x_2 - \sin(x_2)$ as an \emph{evaluation trace} of elementary operations---also called a Wengert list \citep{Wengert1964}. We adopt the three-part notation used by \citet{Griewank2008}, where a function $f: \mathbb{R}^n \to \mathbb{R}^m$ is constructed using intermediate variables $v_i$ such that
\begin{compactitem}
  \item variables $v_{i-n} = x_i,\;i = 1, \dotsc, n$ are the input variables,
  \item variables $v_i\;i = 1, \dotsc, l$ are the working (intermediate) variables, and
  \item variables $y_{m-i} = v_{l-i},\;i = m - 1, \dotsc, 0$ are the output variables.
\end{compactitem}
Figure~\ref{FigureComputationalGraph} shows the given trace of elementary operations represented as a computational graph \citep{Bauer1974}, useful in visualizing dependency relations between intermediate variables.

Evaluation traces form the basis of the AD techniques. An important point to note here is that AD can differentiate not only closed-form expressions in the classical sense, but also algorithms making use of control flow such as branching, loops, recursion, and procedure calls, giving it an important advantage over symbolic differentiation which severely limits such expressivity. This is thanks to the fact that any numeric code will eventually result in a numeric evaluation trace with particular values of the input, intermediate, and output variables, which are the only things one needs to know for computing derivatives using chain rule composition, regardless of the specific control flow path that was taken during execution. Another way of expressing this is that AD is blind with respect to any operation, including control flow statements, which do not directly alter numeric values.

\subsection{Forward Mode}

AD in forward accumulation mode\footnote{Also called \emph{tangent linear} mode.} is the conceptually most simple type. Consider the evaluation trace of the function $f(x_1, x_2) = \ln(x_1) + x_1 x_2 - \sin(x_2)$ given on the left-hand side in Table~\ref{TableForwardADExample} and in graph form in Figure~\ref{FigureComputationalGraph}. For computing the derivative of $f$ with respect to $x_1$, we start by associating with each intermediate variable $v_i$ a derivative
\begin{equation*}
  \dot{v}_i = \frac{\partial v_i}{\partial x_1}\; .
\end{equation*}

Applying the chain rule to each elementary operation in the forward primal trace, we generate the corresponding tangent (derivative) trace, given on the right-hand side in Table~\ref{TableForwardADExample}. Evaluating the primals $v_i$ in lockstep with their corresponding tangents $\dot{v}_i$ gives us the required derivative in the final variable $\dot{v}_5=\frac{\partial y}{\partial x_1}$\;.


\begin{table}
  \centering
  \renewcommand{\arraystretch}{1.2}
  \caption{Forward mode AD example, with $y = f(x_1, x_2) = \ln(x_1) + x_1 x_2 - \sin(x_2)$ evaluated at $(x_1, x_2) = (2, 5)$ and setting $\dot{x}_1 = 1$ to compute $\frac{\partial y}{\partial x_1}$. The original forward evaluation of the primals on the left is augmented by the tangent operations on the right, where each line complements the original directly to its left.}
  \label{TableForwardADExample}
  \begin{minipage}[c]{0.47\textwidth}
    {\footnotesize
    \begin{tabularx}{\textwidth}{p{0.2mm}p{2mm}p{16mm}X}
      \toprule
      \multicolumn{4}{l}{Forward Primal Trace}\\
      \multirow{9}{2mm}{\begin{tikzpicture}\draw[->,>=triangle 60,thick](0,0)--(0,-3.8);\end{tikzpicture}} & $v_{-1}$ & $=x_1$ & $=2$\\
      & $v_0$ & $=x_2$ & $=5$\\
      \cmidrule{2-4}
      & $v_1$ & $=\ln{v_{-1}}$ & $=\ln{2}$\\
      & $v_2$ & $=v_{-1} \times v_0$ & $=2 \times 5$\\
      & $v_3$ & $=\sin{v_0}$ & $=\sin{5}$\\
      & $v_4$ & $=v_1+v_2$ & $=0.693+10$\\
      & $v_5$ & $=v_4-v_3$ & $=10.693+0.959$\\
      \cmidrule{2-4}
      & $y$ & $=v_5$ & $=11.652$\\
      \bottomrule
    \end{tabularx}}
  \end{minipage}
  \begin{minipage}[c]{0.52\textwidth}
    \setlength{\fboxsep}{0pt}\colorbox{gray!20}
    {\footnotesize
    \begin{tabularx}{\textwidth}{p{0.2mm}p{2mm}p{28mm}X}
      \toprule
      \multicolumn{4}{l}{Forward Tangent (Derivative) Trace}\\
      \multirow{9}{2mm}{\begin{tikzpicture}\draw[->,>=triangle 60,thick](0,0)--(0,-3.8);\end{tikzpicture}} & $\dot{v}_{-1}$ & $=\dot{x}_1$ & $=1$\\
      & $\dot{v}_0$ & $=\dot{x}_2$ & $=0$\\
      \cmidrule{2-4}
      & $\dot{v}_1$ & $=\dot{v}_{-1}/v_{-1}$ & $=1/2$\\
      & $\dot{v}_2$ & $=\dot{v}_{-1} \times v_0 + \dot{v}_0 \times v_{-1}$ & $=1 \times 5 + 0 \times 2$\\
      & $\dot{v}_3$ & $=\dot{v}_0 \times \cos{v_0}$ & $=0 \times \cos{5}$\\
      & $\dot{v}_4$ & $=\dot{v}_1+\dot{v}_2$ & $=0.5+5$\\
      & $\dot{v}_5$ & $=\dot{v}_4-\dot{v}_3$ & $=5.5-0$\\
      \cmidrule{2-4}
      & \boldmath$\dot{y}$ & \boldmath$=\dot{v}_5$ & \boldmath$=5.5$\\
      \bottomrule
      \end{tabularx}}
  \end{minipage}
\end{table}

This generalizes naturally to computing the Jacobian of a function $f : \mathbb{R}^n \to \mathbb{R}^m$ with $n$ independent (input) variables $x_i$ and $m$ dependent (output) variables $y_j$. In this case, each forward pass of AD is initialized by setting only one of the variables $\dot{x}_i=1$ and setting the rest to zero (in other words, setting $\dot{\mathbf{x}} = \mathbf{e}_i$, where $\mathbf{e}_i$ is the $i$-th unit vector). A run of the code with specific input values $\mathbf{x}=\mathbf{a}$ then computes
\begin{equation*}
\dot{y}_j = \left.\frac{\partial y_j}{\partial x_i}\right|_{\mathbf{x}=\mathbf{a}},\;j = 1, \dotsc, m\;,
\end{equation*}
giving us one column of the Jacobian matrix
\begin{equation*}
\mathbf{J}_f = \left. \begin{bmatrix}
                    \frac{\partial y_1}{\partial x_1} & \cdots & \frac{\partial y_1}{\partial x_n} \\
                    \vdots & \ddots & \vdots \\
                    \frac{\partial y_m}{\partial x_1} & \cdots & \frac{\partial y_m}{\partial x_n}
                   \end{bmatrix} \right|_{\mathbf{x}\; = \; \mathbf{a}}
\end{equation*}
evaluated at point $\mathbf{a}$. Thus, the full Jacobian can be computed in $n$ evaluations.

Furthermore, forward mode AD provides a very efficient and matrix-free way of computing Jacobian--vector products
\begin{equation}
  \mathbf{J}_f\,\mathbf{r} = \begin{bmatrix}
                    \frac{\partial y_1}{\partial x_1} & \cdots & \frac{\partial y_1}{\partial x_n} \\
                    \vdots & \ddots & \vdots \\
                    \frac{\partial y_m}{\partial x_1} & \cdots & \frac{\partial y_m}{\partial x_n}
                   \end{bmatrix}
                   \begin{bmatrix}
                    r_1 \\
                    \vdots \\
                    r_n
                   \end{bmatrix}\; ,
\label{EquationJacobianVectorProduct}
\end{equation}
simply by initializing with $\dot{\mathbf{x}}=\mathbf{r}$. Thus, we can compute the Jacobian--vector product in just one forward pass. As a special case, when $f: \mathbb{R}^n \to \mathbb{R}$, we can obtain the directional derivative along a given vector $\mathbf{r}$ as a linear combination of the partial derivatives
\begin{equation*}
  \nabla f \cdot \mathbf{r}
\end{equation*}
by starting the AD computation with the values $\dot{\mathbf{x}}=\mathbf{r}$.

Forward mode AD is efficient and straightforward for functions $f: \mathbb{R} \to \mathbb{R}^m$, as all the derivatives $\frac{d y_i}{d x}$ can be computed with just one forward pass. Conversely, in the other extreme of $f: \mathbb{R}^n \to \mathbb{R}$, forward mode AD requires $n$ evaluations to compute the gradient
\begin{equation*}
  \nabla f = \left( \frac{\partial y}{\partial x_1}, \dots, \frac{\partial y}{\partial x_n}\right)\; ,
\end{equation*}
which also corresponds to a $1 \times n$ Jacobian matrix that is built one column at a time with the forward mode in $n$ evaluations.

In general, for cases $f: \mathbb{R}^n \to \mathbb{R}^m$ where $n \gg m$, a different technique is often preferred.
We will describe AD in \emph{reverse accumulation mode} in Section~\ref{sec:reverse-mode}.

\subsubsection{Dual Numbers}
\label{SectionDualNumbers}
Mathematically, forward mode AD (represented by the left- and right-hand sides in Table~\ref{TableForwardADExample}) can be viewed as evaluating a function using dual numbers,\footnote{First introduced by \citet{Clifford1873}, with important uses in linear algebra and physics.} which can be defined as truncated Taylor series of the form
\begin{equation*}
  v + \dot{v}\epsilon \;,
\end{equation*}
where $v, \dot{v} \in \mathbb{R}$ and $\epsilon$ is a nilpotent number such
that $\epsilon^2 = 0$ and $\epsilon \neq 0$. Observe, for example, that
\begin{align*}
  (v + \dot{v}\epsilon) + (u + \dot{u}\epsilon) &= (v + u) + (\dot{v} + \dot{u})\epsilon\\
  (v + \dot{v}\epsilon)(u + \dot{u}\epsilon) &= (vu) + (v\dot{u} + \dot{v}u)\epsilon\;,
\end{align*}
in which the coefficients of $\epsilon$ conveniently mirror symbolic differentiation rules (e.g., Eq.~\ref{EquationMultiplicationRule}). We can utilize this by setting up a regime where
\begin{equation}\label{EquationDualRule}
  f(v + \dot{v}\epsilon) = f(v) + f'(v)\dot{v}\epsilon
\end{equation}
and using dual numbers as data structures for carrying the tangent value together with the primal.\footnote{Just as the complex number written $x + y i$ is represented in the computer as a pair in memory $(x, y)$ whose two slots are reals, the dual number written $x + \dot{x}\epsilon$ is represented as the pair $(x, \dot{x})$. Such pairs are sometimes called Argand pairs \citep[][p107 Eqs.~(157) and (158)]{Hamilton1837}.} The chain rule works as expected on this representation: two applications of Eq.~\ref{EquationDualRule} give
\begin{align*}
  f(g(v + \dot{v}\epsilon)) &= f(g(v) + g'(v)\dot{v}\epsilon)\\
  &= f(g(v)) + f'(g(v))g'(v)\dot{v}\epsilon\;.
\end{align*}
The coefficient of $\epsilon$ on the right-hand side is exactly the derivative of the composition of $f$ and $g$. This means that since we implement elementary operations to respect the invariant Eq.~\ref{EquationDualRule}, all compositions of them will also do so. This, in turn, means that we can extract the derivative of a function by interpreting any non-dual number $v$ as $v + 0 \epsilon$ and evaluating the function in this non-standard way on an initial input with a coefficient $1$ for $\epsilon$:
\begin{align*}
\left.\frac{df(x)}{dx}\right|_{x=v} = \textrm{epsilon-coefficient}(\textrm{dual-version}(f)(v + 1\epsilon))\;.
\end{align*}

This also extends to arbitrary program constructs, since dual numbers, as data types, can be contained in any data structure. As long as a dual number remains in a data structure with no arithmetic operations being performed on it, it will just remain a dual number; and if it is taken out of the data structure and operated on again, then the differentiation will continue.

In practice, a function $f$ coded in a programming language of choice would be fed into an AD tool, which would then augment it with corresponding extra code to handle the dual operations so that the function and its derivative are simultaneously computed. This can be implemented through calls to a specific library, in the form of source code transformation where a given source code will be automatically modified, or through operator overloading, making the process transparent to the user. We discuss these implementation techniques in Section~\ref{SectionImplementations}.

\subsection{Reverse Mode}
\label{sec:reverse-mode}

AD in the reverse accumulation mode\footnote{Also called \emph{adjoint} or \emph{cotangent linear} mode.} corresponds to a generalized backpropagation algorithm, in that it propagates derivatives backward from a given output. This is done by complementing each intermediate variable $v_i$ with an adjoint
\begin{equation*}
  \bar{v}_i = \frac{\partial y_j}{\partial v_i}\; ,
\end{equation*}
which represents the sensitivity of a considered output $y_j$ with respect to changes in $v_i$. In the case of backpropagation, $y$ would be a scalar corresponding to the error $E$ (Figure~\ref{FigureBackpropagation}).

In reverse mode AD, derivatives are computed in the second phase of a two-phase process. In the first phase, the original function code is run \emph{forward}, populating intermediate variables $v_i$ and recording the dependencies in the computational graph through a bookkeeping procedure. In the second phase, derivatives are calculated by propagating adjoints $\bar{v}_i$ in \emph{reverse}, from the outputs to the inputs.

Returning to the example $y = f(x_1, x_2) = \ln(x_1) + x_1 x_2 - \sin(x_2)$, in Table~\ref{TableReverseADExample} we see the adjoint statements on the right-hand side, corresponding to each original elementary operation on the left-hand side. In simple terms, we are interested in computing the contribution $\bar{v}_i = \frac{\partial y}{\partial v_i}$ of the change in each variable $v_i$ to the change in the output $y$. Taking the variable $v_0$ as an example, we see  in Figure~\ref{FigureComputationalGraph} that the only way it can affect $y$ is through affecting $v_2$ and $v_3$, so its contribution to the change in $y$ is given by
\begin{align*}
  \frac{\partial y}{\partial v_0} &= \frac{\partial y}{\partial v_2}\frac{\partial v_2}{\partial v_0} + \frac{\partial y}{\partial v_3}\frac{\partial v_3}{\partial v_0}&
  \text{or}&&
  \bar{v}_0 &= \bar{v}_2\frac{\partial v_2}{\partial v_0} + \bar{v}_3\frac{\partial v_3}{\partial v_0}\;.
\end{align*}

In Table~\ref{TableReverseADExample}, this contribution is computed in two incremental steps
\begin{align*}
  \bar{v}_0 &= \bar{v}_3\frac{\partial v_3}{\partial v_0} &
  \text{and} &&
  \bar{v}_0 &= \bar{v}_0 + \bar{v}_2\frac{\partial v_2}{\partial v_0}\;,
\end{align*}
lined up with the lines in the forward trace from which these expressions originate.

After the forward pass on the left-hand side, we run the reverse pass of the adjoints on the right-hand side, starting with $\bar{v}_5 = \bar{y} = \frac{\partial y}{\partial y} = 1$. In the end we get the derivatives $\frac{\partial y}{\partial x_1} = \bar{x}_1$ and $\frac{\partial y}{\partial x_2} = \bar{x}_2$ in just one reverse pass.

\begin{table}
  \centering
  \renewcommand{\arraystretch}{1.2}
  \caption{Reverse mode AD example, with $y = f(x_1, x_2) = \ln(x_1) + x_1 x_2 - \sin(x_2)$ evaluated at $(x_1, x_2) = (2, 5)$. After the forward evaluation of the primals on the left, the adjoint operations on the right are evaluated in reverse (cf.\ Figure~\ref{FigureBackpropagation}). Note that both $\frac{\partial y}{\partial x_1}$ and $\frac{\partial y}{\partial x_2}$ are computed in the same reverse pass, starting from the adjoint $\bar{v}_5 = \bar{y} = \frac{\partial y}{\partial y} = 1$.}
  \label{TableReverseADExample}
  \begin{minipage}[t]{0.41\textwidth}
    {\footnotesize
    \begin{tabularx}{\textwidth}[t]{p{0.5mm}p{0.8mm}p{18mm}@{}X}
      \toprule
      \multicolumn{4}{l}{Forward Primal Trace}\\
      \multirow{9}{1mm}{\begin{tikzpicture}\draw[->,>=triangle 60,thick](0,0)--(0,-5.4);\end{tikzpicture}} & $v_{-1}$ & $=x_1$ & $=2$\\
      & $v_0$ & $=x_2$ & $=5$\\
      \cmidrule{2-4}
      & $v_1$ & $=\ln{v_{-1}}$ & $=\ln{2}$\vspace{0.25mm}\\
      & $v_2$ & $=v_{-1} \times v_0$ & $=2 \times 5$\vspace{0.25mm}\\
      &\vspace{0.25mm}\\
      & $v_3$ & $=\sin{v_0}$ & $=\sin{5}$\vspace{0.25mm}\\
      & $v_4$ & $=v_1+v_2$ & $=0.693+10$\vspace{0.25mm}\\
      &\vspace{0.25mm}\\
      & $v_5$ & $=v_4-v_3$ & $=10.693+0.959$\vspace{0.25mm}\\
      &\vspace{0.25mm}\\
      \cmidrule{2-4}
      & $y$ & $=v_5$ & $=11.652$\\
      \bottomrule
    \end{tabularx}}\vspace{1mm}
  \end{minipage}
  \begin{minipage}[t]{0.58\textwidth}
    \setlength{\fboxsep}{0pt}\colorbox{gray!20}
    {\footnotesize
    \begin{tabularx}{\textwidth}[t]{p{0.5mm}p{1mm}p{23mm}@{\hspace{1mm}}p{24mm}@{}X}
      \toprule
      \multicolumn{5}{l}{Reverse Adjoint (Derivative) Trace}\\
      \multirow{9}{1mm}{\begin{tikzpicture}\draw[<-,>=triangle 60,thick](0,0)--(0,-5.4);\end{tikzpicture}} & \boldmath$\bar{x}_1$ & \boldmath$=\bar{v}_{-1}$ & & \boldmath$=5.5$\\
      & \boldmath$\bar{x}_2$ & \boldmath$=\bar{v}_0$ & & \boldmath$=1.716$\\
      \cmidrule{2-5}
      & $\bar{v}_{-1}$ & $=\bar{v}_{-1} + \bar{v}_1 \frac{\partial v_1}{\partial v_{-1}}$ & $=\bar{v}_{-1} + \bar{v}_1 / v_{-1}$ & $=5.5$\\
      & $\bar{v}_0$ & $=\bar{v}_0 + \bar{v}_2 \frac{\partial v_2}{\partial v_0}$ & $=\bar{v}_0 + \bar{v}_2 \times v_{-1}$ & $=1.716$\\
      & $\bar{v}_{-1}$ & $=\bar{v}_2 \frac{\partial v_2}{\partial v_{-1}}$ & $=\bar{v}_2 \times v_0$ & $=5$\\
      & $\bar{v}_0$ & $=\bar{v}_3 \frac{\partial v_3}{\partial v_0}$ & $=\bar{v}_3 \times \cos{v_0}$ & $=-0.284$\\
      & $\bar{v}_2$ & $=\bar{v}_4 \frac{\partial v_4}{\partial v_2}$ & $=\bar{v}_4 \times 1$ & $=1$\\
      & $\bar{v}_1$ & $=\bar{v}_4 \frac{\partial v_4}{\partial v_1}$ & $=\bar{v}_4 \times 1$ & $=1$\\
      & $\bar{v}_3$ & $=\bar{v}_5 \frac{\partial v_5}{\partial v_3}$ & $=\bar{v}_5 \times (-1)$ & $=-1$\\
      & $\bar{v}_4$ & $=\bar{v}_5 \frac{\partial v_5}{\partial v_4}$ & $=\bar{v}_5 \times 1$ & $=1$\\
      \cmidrule{2-5}
      & $\bar{v}_5$ & $=\bar{y}$ & $=1$\\
      \bottomrule
      \end{tabularx}}
  \end{minipage}
\end{table}

Compared with the straightforwardness of forward accumulation mode, reverse mode AD can, at first, appear somewhat ``mysterious'' \citep{Dennis1996}. \citet{Griewank2008} argue that this is in part because of the common acquaintance with the chain rule as a mechanistic procedure propagating derivatives forward.

An important advantage of the reverse mode is that it is significantly less costly to evaluate (in terms of operation count) than the forward mode for functions with a large number of inputs. In the extreme case of $f: \mathbb{R}^n \to \mathbb{R}$, only one application of the reverse mode is sufficient to compute the full gradient $\nabla f = \left(\frac{\partial y}{\partial x_1},\dots,\frac{\partial y}{\partial x_n}\right)$, compared with the $n$ passes of the forward mode needed for populating the same. Because machine learning practice principally involves the gradient of a scalar-valued objective with respect to a large number of parameters, this establishes the reverse mode, as opposed to the forward mode, as the mainstay technique in the form of the backpropagation algorithm.

In general, for a function $f: \mathbb{R}^n \to \mathbb{R}^m$, if we denote the operation count to evaluate the original function by $\textrm{ops}(f)$, the time it takes to calculate the $m \times n$ Jacobian by the forward mode is $n\;c\;\textrm{ops}(f)$, whereas the same computation can be done via reverse mode in $m\;c\;\textrm{ops}(f)$, where $c$ is a constant guaranteed to be $c<6$ and typically $c \sim [2,3]$ \citep{Griewank2008}. That is to say, reverse mode AD performs better when $m \ll n$.

Similar to the matrix-free computation of Jacobian--vector products with forward mode (Eq.~\ref{EquationJacobianVectorProduct}), reverse mode can be used for computing the transposed Jacobian--vector product
\begin{equation*}
  \mathbf{J}^{\intercal}_f\,\mathbf{r} = \begin{bmatrix}
                    \frac{\partial y_1}{\partial x_1} & \cdots & \frac{\partial y_m}{\partial x_1} \\
                    \vdots & \ddots & \vdots \\
                    \frac{\partial y_1}{\partial x_n} & \cdots & \frac{\partial y_m}{\partial x_n}
                   \end{bmatrix}
                   \begin{bmatrix}
                    r_1 \\
                    \vdots \\
                    r_m
                   \end{bmatrix}\;,
\end{equation*}
by initializing the reverse phase with $\bar{\mathbf{y}}=\mathbf{r}$.

The advantages of reverse mode AD, however, come with the cost of increased storage requirements growing (in the worst case) in proportion to the number of operations in the evaluated function. It is an active area of research to improve storage requirements in implementations by using advanced methods such as checkpointing strategies and data-flow analysis \citep{Dauvergne2006,siskind2017divide}.

\subsection{Origins of AD and Backpropagation}

Ideas underlying AD date back to the 1950s \citep{Nolan1953,Beda1959}. Forward mode AD as a general method for evaluating partial derivatives was essentially discovered by \citet{Wengert1964}. It was followed by a period of relatively low activity, until interest in the field was revived in the 1980s mostly through the work of \citet{Griewank1989}, also supported by improvements in modern programming languages and the feasibility of an efficient reverse mode AD.

Reverse mode AD and backpropagation have an intertwined history. The essence of the reverse mode, cast in a continuous-time formalism, is the Pontryagin maximum principle \citep{Rozonoer-Pontryagin-1959a, Boltyanskii-Gamkrelidze-Pontryagin-1960a}. This method was understood in the control theory community \citep{Bryson-1962a, Bryson-Ho-1969a} and cast in more formal terms with discrete-time variables topologically sorted in terms of dependency by \citet{Werbos-1974a}. Prior to Werbos, the work by \citet{linnainmaa1970representation,linnainmaa1976taylor} is often cited as the first published description of the reverse mode. \citet{Speelpenning80} subsequently introduced reverse mode AD as we know it, in the sense that he gave the first implementation that was actually automatic, accepting a specification of a computational process written in a general-purpose programming language and automatically performing the reverse mode transformation.

Incidentally, \citet{Hecht1989} cites the work of \citet{Bryson-Ho-1969a} and \citet{Werbos-1974a} as the two earliest known instances of backpropagation. Within the machine learning community, the method has been reinvented several times, such as by \citet{Parker1985}, until it was eventually brought to fame by \citet{rumelhart1986learning} and the Parallel Distributed Processing (PDP) group. The PDP group became aware of Parker's work only after their own discovery; similarly, Werbos' work was not appreciated until it was found by Parker \citep{Hecht1989}. This tells us an interesting story of two highly interconnected research communities that have somehow also managed to stay detached during this foundational period.

For a thorough review of the development of AD, we advise readers to refer to \citet{Rall2006}. Interested readers are highly recommended to read \citet{Griewank2012} for an investigation of the origins of the reverse mode and \citet{schmidhuber2015deep} for the same for backpropagation.

\section{AD and Machine Learning}
\label{SectionDerivativesAndMachineLearning}

In the following, we examine the main uses of derivatives in machine learning and report on a selection of works where general-purpose AD, as opposed to just backpropagation, has been successfully applied in a machine learning context. Areas where AD has seen use include optimization, neural networks, computer vision, natural language processing, and probabilistic inference.

\subsection{Gradient-Based Optimization}

Gradient-based optimization is one of the pillars of machine learning \citep{bottou2016optimization}. Given an objective function $f: \mathbb{R}^n \to \mathbb{R}$, classical gradient descent has the goal of finding (local) minima $\mathbf{w}^* = \argmin_{\mathbf{w}} f(\mathbf{w})$ via updates of the form $\Del \mathbf{w} = -\eta \nabla f$, where $\eta>0$ is a step size. Gradient-based methods make use of the fact that $f$ decreases steepest if one goes in the direction of the negative gradient. The convergence rate of gradient-based methods is usually improved by adaptive step-size techniques that adjust the step size $\eta$ on every iteration \citep{duchi2011adaptive,schaul2013no,kingma2015adam}.

\begin{figure}
  \centering
  \trimbox{0cm -1cm}{\resizebox{0.65\textwidth}{!}{\normalsize\input{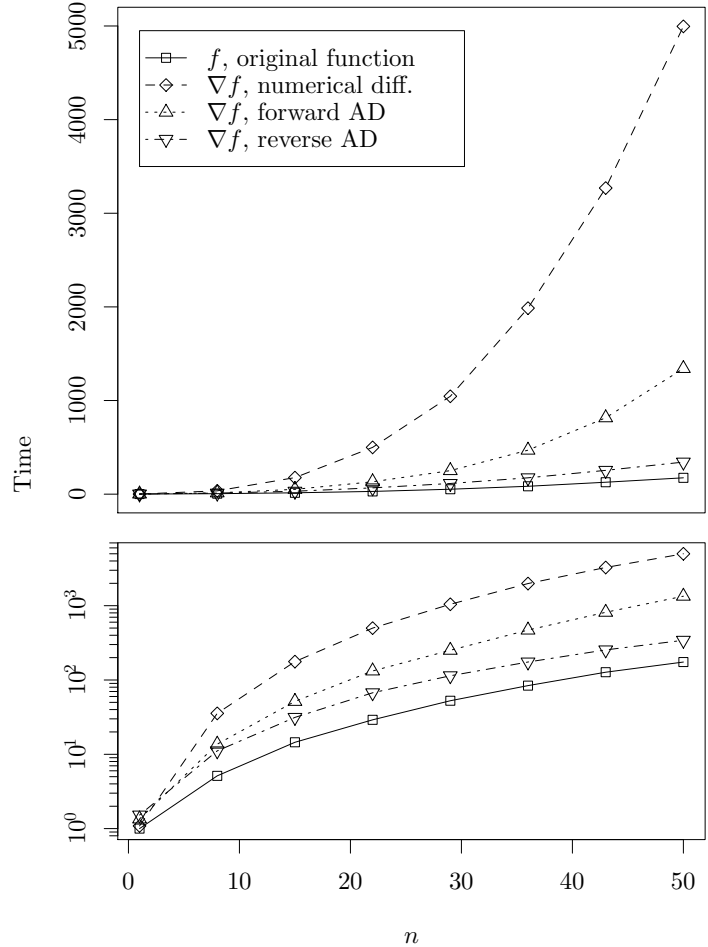}}}
  \caption{Evaluation time of the Helmholtz free energy function of a mixed fluid, based on the Peng-Robinson equation of state \citep{Peng1976}, ${f(\mathbf{x}) = R \, T \sum_{i = 0}^{n} \log \frac{x_i}{1 - \mathbf{b^T} \mathbf{x}} - \frac{\mathbf{x^T} \mathbf{A} \mathbf{x}}{\sqrt{8} \mathbf{b^T} \mathbf{x}} \log \frac{1 + (1 + \sqrt{2}) \mathbf{b^T} \mathbf{x}}{1 + (1 - \sqrt{2}) \mathbf{b^T} \mathbf{x}}}$, where $R$ is the universal gas constant, $T$ is the absolute temperature, $\mathbf{b} \in \mathbb{R}^n$ is a vector of constants, $\mathbf{A} \in \mathbb{R}^{n \times n}$ is a symmetric matrix of constants, and $\mathbf{x} \in \mathbb{R}^n$ is the vector of independent variables describing the system. The plots show the evaluation time of $f$ and the gradient $\nabla f$ with numerical differentiation (central difference), forward mode AD, and reverse mode AD, as a function of the number of variables $n$. Reported times are relative to the evaluation time of $f$ with $n=1$. The lower plot uses logarithmic scale for illustrating the behavior for small $n$. Numerical results are given in Table~\ref{TableHelmholtz}. (Code: {\small\url{http://DiffSharp.github.io/DiffSharp/misc/Benchmarks-h-grad-v0.5.7.fsx}})}
  \label{FigureHelmholtz}
\end{figure}

As we have seen, for large $n$, reverse mode AD provides a highly efficient method for computing gradients.\footnote{See \url{http://DiffSharp.github.io/DiffSharp/examples-gradientdescent.html} for an example of a general-purpose AD-based gradient descent routine using DiffSharp.} Figure~\ref{FigureHelmholtz} and Table~\ref{TableHelmholtz} demonstrate how gradient computation scales differently for forward and reverse mode AD and numerical differentiation, looking at the Helmholtz free energy function that has been used in AD literature for benchmarking gradient calculations \citep{Griewank1989,Griewank2008,griewank2012numerical}.

\begin{table}
  \centering
  \renewcommand{\arraystretch}{1.2}
  \setlength{\tabcolsep}{1.85mm}
  \caption{Evaluation times of the Helmholtz free energy function and its gradient (Figure~\ref{FigureHelmholtz}). Times are given relative to that of the original function with both (1) $n=1$ and (2) $n$ corresponding to each column. (For instance, reverse mode AD with $n=43$ takes approximately twice the time to evaluate relative to the original function with $n=43$.) Times are measured by averaging a thousand runs on a machine with Intel Core i7-4785T 2.20 GHz CPU and 16 GB RAM, using DiffSharp 0.5.7. The evaluation time for the original function with $n=1$ is 0.0023 ms.}
  \label{TableHelmholtz}
  {\small
  \begin{tabularx}{\columnwidth}{@{}p{42.8mm}rrrrrrrr@{}}
    \toprule
    & \multicolumn{4}{l}{$n$, number of variables}\\
    \cmidrule(l){2-9}
    & 1 & 8 & 15 & 22 & 29 & 36 & 43 & 50 \\
    \midrule
    $f$, original\\
    \hspace{2mm} Relative $n=1$ & 1 & 5.12 & 14.51 & 29.11 & 52.58 & 84.00 & 127.33 & 174.44 \\
    $\nabla f$, numerical diff.\\
    \hspace{2mm} Relative $n=1$ & 1.08 & 35.55 & 176.79 & 499.43 & 1045.29 & 1986.70 & 3269.36 & 4995.96 \\
    \hspace{2mm} Relative $n$ in column & 1.08 & 6.93 & 12.17 & 17.15 & 19.87 & 23.64 & 25.67 & 28.63 \\
    $\nabla f$, forward AD\\
    \hspace{2mm} Relative $n=1$ & 1.34 & 13.69 & 51.54 & 132.33 & 251.32 & 469.84 & 815.55 & 1342.07\\
    \hspace{2mm} Relative $n$ in column & 1.34 & 2.66 & 3.55 & 4.54 & 4.77 & 5.59 & 6.40 & 7.69 \\
    $\nabla f$, reverse AD\\
    \hspace{2mm} Relative $n=1$ & 1.52 & 11.12 & 31.37 & 67.27 & 113.99 & 174.62 & 254.15 & 342.33 \\
    \hspace{2mm} Relative $n$ in column & 1.52 & 2.16 & 2.16 & 2.31 & 2.16 & 2.07 & 1.99 & 1.96 \\
    \bottomrule
  \end{tabularx}}
\end{table}

Second-order methods based on Newton's method make use of both the gradient $\nabla f$ and the Hessian $\mathbf{H}_f$, working via updates of the form $\Del \mathbf{w} = -\eta\,\mathbf{H}^{-1}_f \nabla f$ and providing significantly faster convergence \citep{Press2007}. AD provides a way of automatically computing the exact Hessian, enabling succinct and convenient general-purpose implementations.\footnote{See \url{http://DiffSharp.github.io/DiffSharp/examples-newtonsmethod.html} for an implementation of Newton's method with the full Hessian.} Newton's method converges in fewer iterations, but this comes at the cost of having to compute $\mathbf{H}_f$ in each iteration. In large-scale problems, the Hessian is usually replaced by a numerical approximation using first-order updates from gradient evaluations, giving rise to quasi-Newton methods. A highly popular such method is the BFGS\footnote{After Broyden–Fletcher–Goldfarb–Shanno, who independently discovered the method in the 1970s.} algorithm, together with its limited-memory variant L-BFGS \citep{Dennis1996}. On the other hand, Hessians arising in large-scale applications are typically sparse. This sparsity along with symmetry can be readily exploited by AD techniques such as computational graph elimination \citep{Dixon1991}, partial separability \citep{Gay1996}, and matrix coloring and compression \citep{Gebremedhin2009}.

In many cases one does not need the full Hessian but only a Hessian--vector product $\mathbf{H} \mathbf{v}$, which can be computed efficiently using a reverse-on-forward configuration of AD by applying the reverse mode to take the gradient of code produced by the forward mode.\footnote{\citet{Christianson2012ALN} demonstrates that the second derivative can be computed with the same arithmetic operation sequence using forward-on-reverse, reverse-on-forward, and reverse-on-reverse. The taping overheads of these methods may differ in implementation-dependent ways.} Given the function $f:\mathbb{R}^n\to\mathbb{R}$, the evaluation point $\mathbf{x}$, and the vector $\mathbf{v}$, one can accomplish this by first computing the directional derivative $\nabla f \cdot \mathbf{v}$ through the forward mode via setting $\dot{\mathbf{x}}=\mathbf{v}$ and then applying the reverse mode on this result to get $\nabla^2 f \cdot \mathbf{v}=\mathbf{H}_f \mathbf{v}$ \citep{Pearlmutter1994}. This computes $\mathbf{H} \mathbf{v}$ with $O(n)$ complexity, even though $\mathbf{H}$ is a $n \times n$ matrix. Availability of robust AD tools may make more sophisticated optimization methods applicable to large-scale machine-learning problems.  For instance, when fast stochastic Hessian--vector products are available, these can be used as the basis of stochastic Newton's methods \citep{agarwal2016second}, which have the potential to endow stochastic optimization with quadratic convergence.

Another approach for improving the rate of convergence of gradient-based methods is to use gain adaptation methods such as stochastic meta-descent (SMD) \citep{Schraudolph1999}, where stochastic sampling is introduced to avoid local minima and reduce the computational expense. An example using SMD with AD Hessian--vector products is given by \citet{Vishwanathan2006} on conditional random fields (CRF). Similarly, \citet{Schraudolph2003} use Hessian--vector products in their model combining conjugate gradient techniques with stochastic gradient descent.

\subsection{Neural Networks, Deep Learning, Differentiable Programming}

Training of a neural network is an optimization problem with respect to its set of weights, which can in principle be addressed by using any method ranging from evolutionary algorithms \citep{such2017deep} to gradient-based methods such as BFGS \citep{Apostolopoulou2009} or the mainstay stochastic gradient descent \citep{bottou2010large} and its many variants \citep{kingma2015adam,tieleman2012lecture,duchi2011adaptive}. As we have seen, the backpropagation algorithm is only a special case of AD: by applying reverse mode AD to an objective function evaluating a network's error as a function of its weights, we can readily compute the partial derivatives needed for performing weight updates.\footnote{See \url{http://DiffSharp.github.io/DiffSharp/examples-neuralnetworks.html} for an implementation of backpropagation with reverse mode AD.}

The LUSH system \citep{LUSH2002}, and its predecessor SN \citep{bottou-lecun-88}, were the first production systems that targeted efficient neural network simulation while incorporating both a general-purpose programming language and AD. Modern deep learning frameworks provide differentiation capability in one way or another, but the underlying mechanism is not always made clear and confusion abounds regarding the use of the terms ``autodiff'', ``automatic differentiation'', and ``symbolic differentiation'', which are sometimes even used interchangeably. In mainstream frameworks including Theano\footnote{Theano is a computational graph optimizer and compiler with GPU support and it currently handles derivatives in a highly optimized form of symbolic differentiation. The result can be interpreted as a hybrid of symbolic differentiation and reverse mode AD, but Theano does not use the general-purpose reverse accumulation as we describe in this paper. (Personal communication with the authors.)} \citep{Bastien2012}, TensorFlow \citep{abadi2016tensorflow}, Caffe \citep{jia2014caffe}, and CNTK \citep{seide2016cntk} the user first constructs a model as a computational graph using a domain-specific mini language, which then gets interpreted by the framework during execution. This approach has the advantage of enabling optimizations of the computational graph structure (e.g., as in Theano), but the disadvantages of having limited and unintuitive control flow and being difficult to debug. In contrast, the lineage of recent frameworks led by autograd \citep{maclaurin2016modeling}, Chainer \citep{tokui2015chainer}, and PyTorch \citep{paszke2017automatic} provide truly general-purpose reverse mode AD of the type we outline in Section~\ref{SectionPreliminaries}, where the user directly uses the host programming language to define the model as a regular program of the forward computation. This eliminates the need for an interpreter, allows arbitrary control flow statements, and makes debugging simple and intuitive.

Simultaneously with the ongoing adoption of general-purpose AD in machine learning, we are witnessing a modeling-centric terminology emerge within the deep learning community. The terms \emph{define-and-run} and \emph{static computational graph} refer to Theano-like systems where a model is constructed, before execution, as a computational graph structure, which later gets executed with different inputs while remaining fixed. In contrast, the terms \emph{define-by-run} and \emph{dynamic computational graph} refer to the general-purpose AD capability available in newer PyTorch-like systems where a model is a regular program in the host programming language, whose execution dynamically constructs a computational graph on-the-fly that can freely change in each iteration.\footnote{Note that the terms ``static'' and ``dynamic'' here are used in the sense of having a fixed versus non-fixed computational graph topology and not in the sense of data flow architectures.}

\emph{Differentiable programming}\footnote{A term advocated by Christopher Olah (\url{http://colah.github.io/posts/2015-09-NN-Types-FP/}), David Dalrymple (\url{https://www.edge.org/response-detail/26794}), and Yann LeCun (\url{https://www.facebook.com/yann.lecun/posts/10155003011462143}) from a deep learning point of view. Note the difference from \emph{differential} dynamic programming \citep{jacobson1970differential} in optimal control.} is another emerging term referring to the realization that deep learning practice essentially amounts to writing program templates of potential solutions to a problem, which are constructed as differentiable directed graphs assembled from functional blocks whose parameters are learned from examples using gradient-based optimization. Expressed in this paradigm, neural networks are just a class of parameterized differentiable programs composed of building blocks such as feed-forward, convolutional, and recurrent elements. We are increasingly seeing these traditional building blocks freely composed in arbitrary algorithmic structures using control flow, as well as the introduction of novel differentiable architectures such as the neural Turing machine \citep{graves2014neural}, a range of controller--interface abstractions \citep{graves2016hybrid,zaremba2016learning,joulin2015inferring,sukhbaatar2015end}, and differentiable versions of data structures such as stacks, queues, deques \citep{grefenstette2015learning}. Availability of general-purpose AD greatly simplifies the implementation of such architectures by enabling their expression as regular programs that rely on the differentiation infrastructure. Although the differentiable programming perspective on deep learning is new, we note that programming with differentiable functions and having differentiation as a language infrastructure has been the main research subject of the AD community for many decades and realized in a wide range of systems and languages as we shall see in Section~\ref{SectionImplementations}.

There are instances in neural network literature---albeit few---where explicit reference has been made to AD for computing error gradients, such as \citet{Eriksson1998} using AD for large-scale feed-forward networks, and the work by \citet{Yang2008}, where the authors use AD to train a neural-network-based proportional-integral-derivative (PID) controller. Similarly, \citet{Rollins2009} uses reverse mode AD in conjunction with neural networks for the problem of optimal feedback control. Another example is given for continuous time recurrent neural networks (CTRNN) by \citet{AlSeyab2008}, where the authors apply AD for the training of CTRNNs predicting dynamic behavior of nonlinear processes in real time and report significantly reduced training time compared with other methods.

\subsection{Computer Vision}

Since the influential work by \citet{krizhevsky2012imagenet}, computer vision has been dominated by deep learning, specifically, variations of convolutional neural networks \citep{lecun1998gradient}. These models are trained end-to-end, meaning that a mapping from raw input data to corresponding outputs is learned, automatically discovering the representations needed for feature detection in a process called representation learning \citep{bengio2013representation}.

Besides deep learning, an interesting area where AD can be applied to computer vision problems is inverse graphics \citep{horn1977understanding,hinton1997generative}---or analysis-by-synthesis \citep{yildirim2015efficient}---where vision is seen as the inference of parameters for a generative model of a scene. Using gradient-based optimization in inverse graphics requires propagating derivatives through whole image synthesis pipelines including the renderer. \citet{eslami2016attend} use numerical differentiation for this purpose. \citet{loper2014opendr} implement the Open Differentiable Renderer (OpenDR), which is a scene renderer that also supplies derivatives of the image pixels with respect to scene parameters, and demonstrate it in the task of fitting an articulated and deformable 3D model of the human body to image and range data from a Kinect device. Similarly, \citet{kulkarni2015picture} implement a differentiable approximate renderer for the task of inference in probabilistic programs describing scenes.

\citet{srajer2016benchmark} investigate the use of AD for three tasks in computer vision and machine learning, namely bundle adjustment \citep{triggs1999bundle}, Gaussian mixture model fitting, and hand tracking \citep{taylor2014user}, and provide a comprehensive benchmark of various AD tools for the computation of derivatives in these tasks.

\citet{Pock2007} make use of AD in addressing the problems of denoising, segmentation, and recovery of information from stereoscopic image pairs, and note the usefulness of AD in identifying sparsity patterns in large Jacobian and Hessian matrices. In another study, \citet{Grabner2008} use reverse mode AD for GPU-accelerated medical 2D/3D registration, a task involving the alignment of data from different sources such as X-ray images or computed tomography. The authors report a six-fold increase in speed compared with numerical differentiation using center difference (cf.\ our benchmark with the Helmholtz function, Figure~\ref{FigureHelmholtz} and Table~\ref{TableHelmholtz}).

\citet{Barrett2013} present a use of general-purpose AD for the task of video event detection using hidden Markov models (HMMs) and \citet{Dalal2005} object detectors, performing training on a corpus of pre-tracked video using an adaptive step size gradient descent with reverse mode AD. Initially implemented with the R6RS-AD package\footnote{\url{https://github.com/qobi/R6RS-AD}} which provides forward and reverse mode AD in Scheme, the resulting gradient code was later ported to C and highly optimized.\footnote{Personal communication.}

\subsection{Natural Language Processing}

Natural language processing (NLP) constitutes one of the areas where rapid progress is being made by applying deep learning techniques \citep{goldberg2016primer}, with applications in tasks including machine translation \citep{bahdanau2014neural}, language modeling \citep{mikolov2010recurrent}, dependency parsing \citep{chen2014fast}, and question answering \citep{pmlr-v48-kumar16}. Besides deep learning approaches, statistical models in NLP are commonly trained using general purpose or specialized gradient-based methods and mostly remain expensive to train. Improvements in training time can be realized by using online or distributed training algorithms \citep{Gimpel2010}. An example using stochastic gradient descent for NLP is given by \citet{Finkel2008} optimizing conditional random field parsers through an objective function. Related with the work on video event detection in the previous section, \citet{Yu2013} report their work on sentence tracking, representing an instance of grounded language learning paired with computer vision, where the system learns word meanings from short video clips paired with descriptive sentences. The method uses HMMs to represent changes in video frames and meanings of different parts of speech. This work is implemented in C and computes the required gradients using AD through the ADOL-C tool.\footnote{An implementation of the sentence tracker applied to video search using sentence-based queries can be accessed online: \url{http://upplysingaoflun.ecn.purdue.edu/~qobi/cccp/sentence-tracker-video-retrieval.html}}

\subsection{Probabilistic Modeling and Inference}

Inference in probabilistic models can be static, such as compiling a given model to Bayesian networks and using algorithms such as belief propagation for inference; or they can be dynamic, executing a model forward many times and computing statistics on observed values to infer posterior distributions. Markov chain Monte Carlo (MCMC) \citep{Neal1993} methods are often used for dynamic inference, such as the Metropolis--Hastings algorithm based on random sampling \citep{Chib1995}. \citet{Meyer2003} give an example of how AD can be used to speed up Bayesian posterior inference in MCMC, with an application in stochastic volatility. Amortized inference \citep{gershman2014amortized,stuhlmuller2013learning} techniques based on deep learning \citep{le2016inference,ritchie2016deep} work by training neural networks for performing approximate inference in generative models defined as probabilistic programs \citep{gordon2014probabilistic}.

When model parameters are continuous, the Hamiltonian---or, hybrid---Monte Carlo (HMC) algorithm provides improved convergence characteristics avoiding the slow exploration of random sampling, by simulating Hamiltonian dynamics through auxiliary ``momentum variables'' \citep{Duane1987}. The advantages of HMC come at the cost of requiring gradient evaluations of complicated probability models. AD is highly suitable here for complementing probabilistic modeling, because it relieves the user from the manual derivation of gradients for each model.\footnote{See \url{http://diffsharp.github.io/DiffSharp/examples-hamiltonianmontecarlo.html} for an implementation of HMC with reverse mode AD.} For instance, the probabilistic programming language Stan \citep{carpenter2016stan} implements automatic Bayesian inference based on HMC and the No-U-Turn sampler (NUTS) \citep{Hoffman2014} and uses reverse mode AD for the calculation of gradients for both HMC and NUTS \citep{carpenter2015stan}. Similarly, \citet{Wingate2011} demonstrate the use of AD as a non-standard interpretation of probabilistic programs enabling efficient inference algorithms. \citet{kucukelbir2017automatic} present an AD-based method for deriving variational inference (VI) algorithms.

PyMC3 \citep{salvatier2016probabilistic} allows fitting of Bayesian models using MCMC and VI, for which it uses gradients supplied by Theano. Edward \citep{tran2016edward} is a library for deep probabilistic modeling, inference, and criticism \citep{tran2017deep} that supports VI using TensorFlow. Availability of general-purpose AD in this area has enabled new libraries such as Pyro\footnote{\url{http://pyro.ai/}} and ProbTorch \citep{siddharth2017learning} for deep \emph{universal} probabilistic programming with support for recursion and control flow, relying, in both instances, on VI using gradients supplied by PyTorch's reverse mode AD infrastructure.

When working with probabilistic models, one often needs to backpropagate derivatives through sampling operations of random variables in order to achieve stochastic optimization of model parameters. The score-function estimator, or REINFORCE \citep{williams1992simple}, method provides a generally applicable unbiased gradient estimate, albeit with high variance. When working with continuous random variables, one can substitute a random variable by a deterministic and differentiable transformation of a simpler random variable, a method known as the ``reparameterization trick'' \citep{williams1992simple,kingma2014auto,rezende2014stochastic}. For discrete variables, the REBAR \citep{tucker2017rebar} method provides a lower-variance unbiased gradient estimator by using continuous relaxation. A generalization of REBAR called RELAX \citep{grathwohl2017backpropagation} works by learning a free-form control variate parameterized by a neural network and is applicable in both discrete and continuous settings.

\section{Implementations}
\label{SectionImplementations}

It is useful to have an understanding of the different ways in which AD can be implemented. Here we cover major implementation strategies and provide a survey of existing tools.

A principal consideration in any AD implementation is the performance overhead introduced by the AD arithmetic and bookkeeping. In terms of computational complexity, AD guarantees that the amount of arithmetic goes up by no more than a small constant factor \citep{Griewank2008}. On the other hand, managing this arithmetic can introduce a significant overhead if done carelessly. For instance, naïvely allocating data structures for holding dual numbers will involve memory access and allocation for every arithmetic operation, which are usually more expensive than arithmetic operations on modern computers.\footnote{The implementation of forward mode in Julia \citep{revels2016forward} attempts to avoid this, and some current compilers can avoid this expense by unboxing dual numbers \citep{leroy1997effectiveness, jones1993glasgow, jones1991unboxed, siskind2016efficient}. This method is also used to reduce the memory-access overhead in the implementations of forward mode in Stalingrad and the Haskell \emph{ad} library.} Likewise, using operator overloading may introduce method dispatches with attendant costs, which, compared to raw numerical computation of the original function, can easily amount to a slowdown of an order of magnitude.\footnote{Flow analysis \citep{shivers1991control} and/or partial evaluation \citep{jones1993partial}, together with tag stripping \citep{appel1989runtime, peterson1989untagged}, can remove this method dispatch. These, together with unboxing, can often make it possible to completely eliminate the memory access, memory allocation, memory reclamation, and method dispatch overhead of dual numbers \citep{siskind2016efficient}.}

Another major issue is the risk of hitting a class of bugs called ``perturbation confusion'' \citep{SiskindPearlmutter2005a,manzyuk2012confusion}. This essentially means that if two ongoing differentiations affect the same piece of code, the two formal epsilons they introduce (Section~\ref{SectionDualNumbers}) need to be kept distinct. It is very easy to have bugs---particularly in performance-oriented AD implementations---that confuse these in various ways. Such situations can also arise when AD is nested, that is, derivatives are computed for functions that internally compute derivatives.

Translation of mathematics into computer code often requires attention to numeric issues.  For instance, the mathematical expressions $\log(1 + x)$ or $\sqrt{x^2+y^2+z^2}$ or $\tan^{-1}(y/x)$ should not be naïvely translated, but rather expressed as \texttt{log1p(x)}, \texttt{hypot(x,hypot(y,z))}, and \texttt{atan2(y,x)}.  In machine learning, the most prominent example of this is probably the so-called log-sum-exp trick to improve the numerics of calculations of the form $\log\sum_i \exp x_i$.  AD is not immune to such numeric considerations. For example, code calculating $E=\sum_i E_i$, processed by AD, will calculate $\nabla_w E=\sum_i\nabla_w E_i$. If the system is seeking a local minimum of $E$ then $\nabla_w E = \sum_i\nabla_w E_i \rightarrow_t 0$, and naïvely adding a set of large numbers whose sum is near zero is numerically fraught.  This is to say that AD is not immune to the perils of floating point arithmetic, and can sometimes introduce numeric issues which were not present in the primal calculation.  Issues of numeric analysis are outside our present scope, but there is a robust literature on the numerics of AD (e.g., \citet{griewank2012numerical}) involving using subgradients to allow optimization to proceed despite non-differentiability of the objective, appropriate subgradients and approximations for functions like $\lvert \cdot \rvert$ and $\lVert\cdot\rVert_2$ and $\sqrt{\cdot}$ near zero, and a spate of related issues.

One should also be cautious about approximated functions and AD \citep{sirkes-tziperman-1997a}. In this case, if one has a procedure \emph{approximating} an ideal function, AD always gives the derivative of the procedure that was actually programmed, which may not be a good approximation of the derivative of the ideal function that the procedure was approximating. For instance, consider $e^x$ computed by a piecewise-rational approximation routine. Using AD on this routine would produce an approximated derivative in which each piece of the piecewise formula will get differentiated. Even if this would remain an approximation of the derivative of $e^x$, we know that $\frac{de^x}{dx} = e^x$ and the original approximation itself was already a better approximation for the derivative of $e^x$.\footnote{In modern systems this is not an issue, because $e^x$ is a primitive implemented in hardware.} Users of AD implementations must be therefore cautious to \emph{approximate the derivative, not differentiate the approximation}. This would require explicitly approximating a known derivative, in cases where a mathematical function can only be computed approximately but has a well-defined mathematical derivative.

We note that there are similarities as well as differences between machine learning workloads and those studied in the traditional AD literature \citep{baydin2016tricks}. Deep learning systems are generally compute-bound and spend a considerable amount of computation time in highly-optimized numerical kernels for matrix operations \citep{hadjis2015caffe,chetlur2014cudnn}. This is a situation which is arguably amenable to operator-overloading-based AD implementations on high-level operations, as is commonly found in current machine learning frameworks. In contrast, numerical simulation workloads in traditional AD applications can be bandwidth-bound, making source code transformation and compiler optimization approaches more relevant. Another difference worth noting is that whereas high numerical precision is desirable in traditional application domains of AD such as computational fluid dynamics \citep{cohen2009fast}, in deep learning lower-precision is sufficient and even desirable in improving computational efficiency, thanks to the error resiliency of neural networks \citep{gupta2015deep,courbariaux2015binaryconnect}.

There are instances in recent literature where implementation-related experience from the AD field has been put to use in machine learning settings. One particular area of recent interest is implicit and iterative AD techniques \citep{Griewank2008}, which has found use in work incorporating constrained optimization within deep learning \citep{amos2017optnet} and probabilistic graphical models and neural networks \citep{johnson2016composing}. Another example is checkpointing strategies \citep{Dauvergne2006,siskind2017divide}, which allow balancing of application-specific trade-offs between time and space complexities of reverse mode AD by not storing the full tape of intermediate variables in memory and reconstructing these as needed by re-running parts of the forward computation from intermediate checkpoints. This is highly relevant in deep learning workloads running on GPUs with limited memory budgets. A recent example in this area is the work by \citet{gruslys2016memory}, where the authors construct a checkpointing variety of the backpropagation through time (BPTT) algorithm for recurrent neural networks and demonstrate it saving up to 95\% memory usage at the cost of a 33\% increase in computation time in one instance.

In Table~\ref{TableADImplementations} we present a review of notable general-purpose AD implementations.\footnote{Also see the website \url{http://www.autodiff.org/} for a list of tools maintained by the AD community.} A thorough taxonomy of implementation techniques was introduced by \citet{Juedes1991}, which was later revisited by \citet{Bischof2008} and simplified into \emph{elemental}, \emph{operator overloading}, \emph{compiler-based}, and \emph{hybrid} methods. We adopt a similar classification for the following part of this section.


\addtolength{\tabcolsep}{-3pt}
\begin{sidewaystable}
  \centering
  \renewcommand{\arraystretch}{1.5}
  \caption{Survey of AD implementations. Tools developed primarily for machine learning are highlighted in bold.}
  \label{TableADImplementations}
  {\tiny
  \begin{tabularx}{\textwidth}{@{}p{12mm}p{20mm}p{5mm}p{6mm}p{62mm}p{38mm}p{65mm}@{}}
    \toprule
    Language & Tool & Type & Mode & Institution / Project & Reference & URL\\
    \midrule
    AMPL & AMPL & INT & F, R & Bell Laboratories & \citet{Fourer2002} & \tiny\url{http://www.ampl.com/}\\
    C, C++ & ADIC & ST & F, R & Argonne National Laboratory & \citet{Bischof1997} & \tiny\url{http://www.mcs.anl.gov/research/projects/adic/}\\
    & ADOL-C & OO & F, R & Computational Infrastructure for Operations Research & \citet{Walther2012} & \tiny\url{https://projects.coin-or.org/ADOL-C}\\
    C++ & Ceres Solver & LIB & F & Google & & \tiny\url{http://ceres-solver.org/}\\
    & CppAD & OO & F, R & Computational Infrastructure for Operations Research & \citet{Bell2008} & \tiny\url{http://www.coin-or.org/CppAD/}\\
    & FADBAD++ & OO & F, R & Technical University of Denmark & \citet{Bendtsen1996} & \tiny\url{http://www.fadbad.com/fadbad.html}\\
    & Mxyzptlk & OO & F & Fermi National Accelerator Laboratory & \citet{Ostiguy2007} & \\
    C\# & AutoDiff & LIB & R & George Mason Univ., Dept. of Computer Science & \citet{Shtof2013} & \tiny\url{http://autodiff.codeplex.com/}\\
    F\#, C\# & \textbf{DiffSharp} & OO & F, R & Maynooth University, Microsoft Research Cambridge & \citet{baydin2016diffsharp} & \tiny\url{http://diffsharp.github.io}\\
    Fortran & ADIFOR & ST & F, R & Argonne National Laboratory & \citet{Bischof1996} & \tiny\url{http://www.mcs.anl.gov/research/projects/adifor/}\\
    & NAGWare & COM & F, R & Numerical Algorithms Group & \citet{Naumann2005} & \tiny\url{http://www.nag.co.uk/nagware/Research/ad_overview.asp}\\
    & TAMC & ST & R & Max Planck Institute for Meteorology & \citet{Giering1998} & \tiny\url{http://autodiff.com/tamc/}\\
    Fortran, C & COSY & INT & F & Michigan State Univ., Biomedical and Physical Sci. & \citet{Berz1996} & \tiny\url{http://www.bt.pa.msu.edu/index_cosy.htm}\\
    & Tapenade & ST & F, R & INRIA Sophia-Antipolis & \citet{Hascoet2013} & \tiny\url{http://www-sop.inria.fr/tropics/tapenade.html}\\
    Haskell & ad & OO & F, R & Haskell package & & \tiny\url{http://hackage.haskell.org/package/ad}\\
    Java & ADiJaC & ST & F, R & University Politehnica of Bucharest & \citet{slusanschi2016adijac} & \tiny\url{http://adijac.cs.pub.ro}\\
    & Deriva & LIB & R & Java \& Clojure library & & \tiny\url{https://github.com/lambder/Deriva}\\
    Julia & JuliaDiff & OO & F, R & Julia packages & \citet{RevelsLubinPapamarkou2016} & \tiny\url{http://www.juliadiff.org/}\\
    Lua & \textbf{torch-autograd} & OO & R & Twitter Cortex & & \tiny\url{https://github.com/twitter/torch-autograd}\\
    MATLAB & ADiMat & ST & F, R & Technical University of Darmstadt, Scientific Comp. & \citet{Willkomm2013} & \tiny\url{http://adimat.sc.informatik.tu-darmstadt.de/}\\
    & INTLab & OO & F & Hamburg Univ. of Technology, Inst. for Reliable Comp. & \citet{Rump1999} & \tiny\url{http://www.ti3.tu-harburg.de/rump/intlab/}\\
    & TOMLAB/MAD & OO & F & Cranfield University \& Tomlab Optimization Inc. & \citet{Forth2006} & \tiny\url{http://tomlab.biz/products/mad}\\
    Python & ad & OO & R & Python package & & \tiny\url{https://pypi.python.org/pypi/ad}\\
    & \textbf{autograd} & OO & F, R & Harvard Intelligent Probabilistic Systems Group & \citet{maclaurin2016modeling} & \tiny\url{https://github.com/HIPS/autograd}\\
    & \textbf{Chainer} & OO & R & Preferred Networks & \citet{tokui2015chainer} & \tiny\url{https://chainer.org/}\\
    & \textbf{PyTorch} & OO & R & PyTorch core team & \citet{paszke2017automatic} & \tiny\url{http://pytorch.org/}\\
    & \textbf{Tangent} & ST & F, R & Google Brain & \citet{van2017tangent} & \tiny\url{https://github.com/google/tangent}\\
    Scheme & R6RS-AD & OO & F, R & Purdue Univ., School of Electrical and Computer Eng. & & \tiny\url{https://github.com/qobi/R6RS-AD}\\
    & Scmutils & OO & F & MIT Computer Science and Artificial Intelligence Lab. & \citet{Sussman2001} & \tiny\url{http://groups.csail.mit.edu/mac/users/gjs/6946/refman.txt}\\
    & Stalingrad & COM & F, R & Purdue Univ., School of Electrical and Computer Eng. & \citet{pearlmutter2008reverse} & \tiny\url{http://www.bcl.hamilton.ie/~qobi/stalingrad/}\\
    \bottomrule
    \addlinespace
    \multicolumn{7}{l}{F: Forward, R: Reverse; COM: Compiler, INT: Interpreter, LIB: Library, OO: Operator overloading, ST: Source transformation}
  \end{tabularx}
  }
\end{sidewaystable}
\addtolength{\tabcolsep}{3pt}

\subsection{Elemental Libraries}

These implementations form the most basic category and work by replacing mathematical operations with calls to an AD-enabled library. Methods exposed by the library are then used in function definitions, meaning that the decomposition of any function into elementary operations is done manually when writing the code.

The approach has been utilized since the early days of AD, with prototypical examples being the WCOMP and UCOMP packages of \citet{Lawson1971}, the APL package of \citet{Neidinger1989}, and the work by \citet{Hinkins1994}. Likewise, \citet{Hill1992} formulate their implementation of AD in MATLAB using elemental methods.

Elemental libraries still constitute the simplest strategy to implement AD for languages without operator overloading.

\subsection{Compilers and Source Code Transformation}

These implementations provide extensions to programming languages that automate the decomposition of algorithms into AD-enabled elementary operations. They are typically executed as preprocessors\footnote{Preprocessors transform program source code before it is given as an input to a compiler.} to transform the input in the extended language into the original language.

Classical instances of source code transformation include the Fortran preprocessors GRESS \citep{Horwedel1988} and PADRE2 \citep{Kubo1990}, which transform AD-enabled variants of Fortran into standard Fortran 77 before compiling. Similarly, the ADIFOR tool \citep{Bischof1996}, given a Fortran source code, generates an augmented code in which all specified partial derivatives are computed in addition to the original result. For procedures coded in ANSI C, the ADIC tool \citep{Bischof1997} implements AD as a source code transformation after the specification of dependent and independent variables. A recent and popular tool also utilizing this approach is Tapenade \citep{Pascual2008,Hascoet2013}, implementing forward and reverse mode AD for Fortran and C programs. Tapenade itself is implemented in Java and can be run locally or as an online service.\footnote{\url{http://www-tapenade.inria.fr:8080/tapenade/index.jsp}}

In addition to language extensions through source code transformation, there are implementations introducing new languages with tightly integrated AD capabilities through special-purpose compilers or interpreters. Some of the earliest AD tools such as SLANG \citep{Adamson1969} and PROSE \citep{Pfeiffer1987} belong to this category. The NAGWare Fortran 95 compiler \citep{Naumann2005} is a more recent example, where the use of AD-related extensions triggers automatic generation of derivative code at compile time.

As an example of interpreter-based implementation, the algebraic modeling language AMPL \citep{Fourer2002} enables objectives and constraints to be expressed in mathematical notation, from which the system deduces active variables and arranges the necessary AD computations. Other examples in this category include the FM/FAD package \citep{Mazourik1991}, based on the Algol-like DIFALG language, and the object-oriented COSY language \citep{Berz1996} similar to Pascal.

The Stalingrad compiler \citep{pearlmutter2008reverse,Siskind2008}, working on the Scheme-based AD-aware VLAD language, also falls under this category. The newer DVL compiler\footnote{\url{https://github.com/axch/dysvunctional-language}} is based on Stalingrad and uses a reimplementation of portions of the VLAD language.

Motivated by machine learning applications, the Tangent library \citep{van2017tangent} implements AD using source code transformation, and accepts numeric functions written in a syntactic subset of Python and Numpy.

\subsection{Operator Overloading}

In modern programming languages with polymorphic features, operator overloading provides the most straightforward way of implementing AD, exploiting the capability of redefining elementary operation semantics.

A popular tool implemented with operator overloading in C++ is ADOL-C \citep{Walther2012}. ADOL-C requires the use of AD-enabled types for variables, and records arithmetic operations on variables in tape data structures, which can subsequently be ``played back'' during reverse mode AD computations. The Mxyzptlk package \citep{Michelotti1990} is another example for C++ capable of computing arbitrary-order partial derivatives via forward propagation. The FADBAD++ library \citep{Bendtsen1996} implements AD for C++ using templates and operator overloading. For Python, the \emph{ad} package\footnote{\url{http://pythonhosted.org/ad/}} uses operator overloading to compute first- and second-order derivatives, while the newer autograd package\footnote{\url{https://github.com/HIPS/autograd}} provides forward and reverse mode AD with support for higher-order derivatives.

For functional languages, examples include R6RS-AD\footnote{\url{https://github.com/NUIM-BCL/R6RS-AD}} and the AD routines within the Scmutils library\footnote{\url{http://groups.csail.mit.edu/mac/users/gjs/6946/refman.txt}} for Scheme, the \emph{ad} library\footnote{\url{http://hackage.haskell.org/package/ad}} for Haskell, and DiffSharp\footnote{\url{http://diffsharp.github.io}} for F\# and C\#.


\section{Conclusions}
\label{SectionConclusions}
Backpropagation and gradient-based optimization are behind virtually all recent successes in machine learning, yielding state-of-the-art results in computer vision, speech recognition and synthesis, and machine translation. We expect these techniques to remain at the core of machine learning for the foreseeable future. Research in the field involves a rapid prototyping and development cycle for testing new models and ideas, using a collection of increasingly higher-quality machine learning frameworks. These frameworks are in the process of transition from coarse-grained (module level) backpropagation towards fine-grained, general-purpose AD, allowing models to be implemented as regular programs in general-purpose programming languages with differentiation as an integral part of the infrastructure. We strongly believe that general-purpose AD is the future of gradient-based machine learning and we expect it to become an indispensable tool in the machine learning toolbox.

It is an exciting time for working at the intersection of AD and machine learning, and there are many opportunities for bringing advanced techniques and expertise from AD literature to bear on machine learning problems. Techniques that have been developed by the AD community such as tape reduction and elimination \citep{naumann2004optimal}, fixed-point iterations \citep{christianson1994reverse}, utilizing sparsity by matrix coloring \citep{Gebremedhin2009,gebremedhin2013colpack}, and reverse AD checkpointing \citep{Dauvergne2006} are just a few examples that can find potential use in machine learning for increasing performance, improving convergence of optimization, using hardware more efficiently, and even enabling new types of machine learning models to be implemented. Similarly, exciting new AD modes like direct propagation of the inverse Jacobian \citep{srinivasan-todorov-2015a} have emerged from the machine learning community, but have yet to be examined and formalized by the AD community.

An important direction for future work is to make use of nested AD techniques in machine learning, allowing differentiation to be nested arbitrarily deep with referential transparency \citep{Siskind2008b,pearlmutter2008reverse}. Nested AD is highly relevant in hyperparameter optimization as it can effortlessly provide exact hypergradients, that is, derivatives of a training objective with respect to the hyperparameters of an optimization routine \citep{Maclaurin2015,baydin2017online}. Potential applications include Bayesian model selection \citep{Rasmussen2006} and gradient-based tuning of Hamiltonian Monte Carlo step sizes and mass matrices \citep{Salimans2014}. Besides hyperparameters, models internally using higher-order derivatives constitute a straightforward usage case for nested AD. The Riemannian manifold Langevin and Hamiltonian Monte Carlo methods \citep{Girolami2011} use higher-order derivative information to more closely track the information geometry of the sampled distribution for faster convergence and exploration. In neural networks, it is very natural to use nested derivatives in defining objective functions that take input transformations into account, such as the Tangent Prop method \citep{Simard1998} for imposing invariance under a set of chosen transformations.


\acks{We thank the anonymous reviewers whose comments helped improve this manuscript. This work was supported, in part, by Science Foundation Ireland grant 09/IN.1/I2637, by the Army Research Laboratory, accomplished under Cooperative Agreement Number W911NF-10-2-0060, by the National Science Foundation under Grants 1522954-IIS and 1734938-IIS, and by the Intelligence Advanced Research Projects Activity (IARPA) via Department of Interior/Interior Business Center (DOI/IBC) contract number D17PC00341. Any opinions, findings, views, and conclusions or recommendations expressed in this material are those of the authors and do not necessarily reflect the views, official policies, or endorsements, either expressed or implied, of the sponsors. The U.S. Government is authorized to reproduce and distribute reprints for Government purposes, notwithstanding any copyright notation herein.
}

%
%



\vskip 0.2in
\bibliography{17-468}

\end{document}